\pdfoutput=1
\documentclass[conference]{IEEEtran}
\IEEEoverridecommandlockouts
\usepackage{cite}
\usepackage{hyperref}
\hypersetup{hidelinks,bookmarks=false}
\usepackage{amsmath,amssymb,amsfonts}
\usepackage{algorithmic}
\usepackage{graphicx}
\usepackage{textcomp}
\usepackage{xcolor}
\usepackage{siunitx}
\usepackage{subcaption}
\usepackage{float}
\newcommand{\colr}{\textcolor{black}}
\newcommand{\colb}{\textcolor{black}}
\newcommand{\colblue}
{\textcolor{black}}
\usepackage[compact]{titlesec}
\titlespacing{\section}{0pt}{1ex}{1ex}
\titlespacing{\subsection}{0pt}{1ex}{1ex}
\titlespacing{\subsubsection}{0pt}{0.5ex}{0.5ex}

\IEEEoverridecommandlockouts
\makeatletter
\def\ps@IEEEtitlepagestyle{
  \def\@oddfoot{\mycopyrightnotice}
  \def\@evenfoot{}
}
\def\mycopyrightnotice{
  {\footnotesize
  \begin{minipage}{\textwidth}
  \copyright~2023 IEEE.~Personal use of this material is permitted. Permission from IEEE must be obtained for all other uses, in any current or future media, including reprinting/republishing this material for advertising or promotional purposes, creating new collective works, for resale or redistribution to servers or lists, or reuse of any copyrighted component of this work in other works.
  \end{minipage}
  }
}
\def\BibTeX{{\rm B\kern-.05em{\sc i\kern-.025em b}\kern-.08em
    T\kern-.1667em\lower.7ex\hbox{E}\kern-.125emX}}
\begin{document}

\title{Can 5G NR Sidelink communications support wireless augmented reality?\\

}


\author{
\IEEEauthorblockN{Ashutosh~Srivastava\IEEEauthorrefmark{1}\IEEEauthorrefmark{2}, Qing~Zhao\IEEEauthorrefmark{1}, Yi~Lu\IEEEauthorrefmark{1}, Ping~Wang\IEEEauthorrefmark{1}, Qi~Qu\IEEEauthorrefmark{1}, Zhu~Ji\IEEEauthorrefmark{1}, Yee~Sin~Chan\IEEEauthorrefmark{1}, Shivendra~S.~Panwar\IEEEauthorrefmark{2}}
\IEEEauthorblockA{
\IEEEauthorrefmark{1}{Meta Platforms, Inc., USA, \{lulouis, pingwang, qqu, zhuji\}@meta.com}\\
\IEEEauthorrefmark{2}{New York University, New York, USA, \{as12738, sp1832\}}@nyu.edu \vspace{-2mm}}}
\maketitle

\begin{abstract}
Smart glasses that support augmented reality (AR) have the potential to become the consumer’s primary medium of connecting to the future internet. For the best quality of user experience, AR glasses must have a small form factor and long battery life, while satisfying the data rate and latency requirements of AR applications. To extend the AR glasses' battery life, the computation and processing involved in AR may be offloaded to a companion device, such as a smartphone, through a  wireless connection. Sidelink (SL), i.e., the D2D communication interface of 5G NR, is a potential candidate for this wireless link. In this paper, we use system-level simulations to analyze the feasibility of NR SL for supporting AR. Our simulator incorporates the PHY layer structure and MAC layer resource scheduling of 3GPP SL, standard 3GPP channel models, and MCS configurations. Our results suggest that the current SL standard specifications are insufficient for high-end AR use cases with heavy interaction but can support simpler previews and file transfers. We further propose two enhancements to SL resource allocation, which have the potential to offer significant performance improvements for AR applications.
\end{abstract}

\begin{IEEEkeywords}
Augmented reality, Sidelink, 5G NR, D2D, 3GPP, Resource allocation
\end{IEEEkeywords}

\section{Introduction}\label{intro}

Augmented reality (AR) involves overlaying virtual information on top of a user's live 3D view of the real world. High-quality AR experiences can be delivered through AR glasses. AR glasses will enable applications that traditional handheld devices, like smartphones, cannot realize. These include many flavors of use cases such as 3D holographic calling, real-time sharing and overlay of virtual objects, real-time frictionless internet browsing, and messaging~\cite{3gpp.26.928}. AR applications require advanced processing and computation to \colb{estimate the user's location under 6DoF motion, identify objects in their field-of-view (FoV), and render the virtual content on top of their real-world view.} The AR glasses \colb{may} also need to maintain a reliable, low latency, and high-speed internet connection to support some use cases. However, for the best user experience, the wearable AR glasses must have a small form factor and long battery life, which limits their computation capacity. In such a scenario, we may need a companion device tethered to the glasses to perform the bulk of the processing. The companion device, such as a smartphone with cellular connectivity, serves as the bridge between the AR glasses and the internet. 



A wired connection between the AR glasses and the companion device provides high-speed connectivity but restricts the user's mobility and degrades their comfort level. For a seamless experience, the AR glasses must have a wireless connection. This device-to-device (D2D) wireless connection between the glasses and the companion will be a critical component of the AR ecosystem. In many scenarios, we need low-latency D2D transfers to ensure a sufficient latency budget for the other components of the AR processing pipeline. The D2D link must also operate within the thermal constraints of the AR glasses, requiring \colb{sufficient duty cycling.} Possible candidate technologies for this D2D wireless link are: Bluetooth low energy (BLE), Wi-Fi, and cellular D2D, i.e., sidelink (SL)~\cite{lien20203gpp}.


\begin{figure}
    \centering
    \includegraphics[width=\columnwidth]{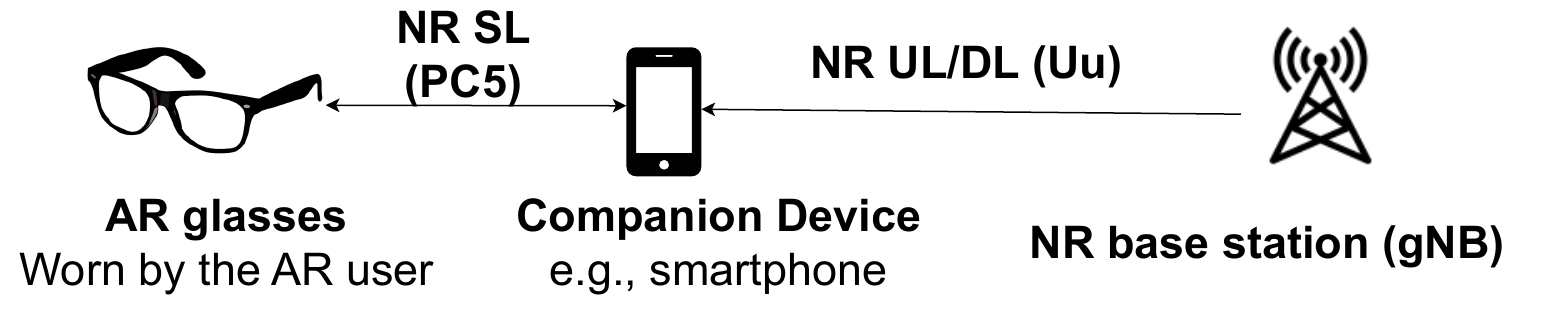}
    \caption{AR user's glasses connected to the cellular network via a wireless SL (D2D) connection to a companion device}
    \label{fig:model}
    \vspace{-0.5cm}
\end{figure}

In this work, we study the feasibility of using SL to support the D2D connection in AR. The model for this is shown in Fig.~\ref{fig:model}. The 3rd Generation Partnership Project (3GPP) first standardized D2D communication for proximity services in its Release 12 and has since updated it in later releases, including the introduction of SL as the D2D interface of 5G NR in 3GPP Rel. 16 (2020). Although, as of Rel. 17, SL is primarily designed for V2X use cases, it will likely be enhanced for AR/VR in the future (Rel. 18 and beyond). This paper conducts an in-depth system-level evaluation of the feasibility of 5G NR SL for supporting AR use cases, motivated by the increasing interest of significant industry players in SL technology~\cite{qualcomm_sidelink}. We believe we are the first to do such an analysis, which differs from previous simulation-based studies on LTE C-V2X and 5G NR V2X~\cite{wang2019system,bazzi2021design, todisco2021performance,ali20213gpp,saad2021advancements}, that primarily focused on V2X deployments and traffic conditions~\cite{3gpp.37.885}. While some recent works~\cite{saafi2021cellular,petrov2022extended} discuss the potential of SL for XR, they do not present a complete performance evaluation.

\colblue{The \textbf{key contributions} of this work are as follows: \\ We develop an abstracted use case framework that involves a connection between the AR glasses and the companion device. We provide a semi-quantitative analysis of the supportability of a spectrum of AR use cases with SL. For this purpose,} we design a system-level simulation setup in MATLAB, which models the PHY layer structure and MAC layer protocols of NR SL. We evaluate SL performance in deployments representative of typical AR use case conditions \colb{with reasonable assumptions}. 3GPP standard channel models, modulation and coding schemes (MCS), and antenna configurations are used along with the link layer abstraction model from \cite{lagen2020new} to ensure that our setup aligns with realistic wireless channel conditions. We identify key limitations that hinder the feasibility of a SL-based D2D connection for AR glasses. Finally, we propose two promising solutions for SL resource allocation: multiple active reservations and full-duplex sensing, which can significantly enhance SL performance for AR use cases. 

The rest of the paper is organized as follows: In section~\ref{introSL}, we provide a basic introduction to 5G NR SL, focusing on its PHY structure and MAC layer resource scheduling. The methodology of our analysis is described in Sec.~\ref{methodology}. We then present the results of this study in Sec.~\ref{results}. Finally,  Sec.~\ref{conclude} concludes our study and presents future research directions.


\section{5G NR Sidelink: Introduction}\label{introSL}
We start with a basic introduction to 5G NR SL. This will serve as the background for the subsequent sections. 
\subsection{SL transmissions at the PHY layer}
NR SL supports unicast, multicast, and broadcast transmissions. The communication between the AR glasses and the companion device is unicast. We assume that SL operates in the sub-7 GHz frequency bands (FR1).\footnote{Optimized SL design for mmWave frequencies (FR2 bands) is yet to be addressed by 3GPP, with only some ongoing research~\cite{mura2022spatial,srivastava2022overcoming}. Hence, this work does not cover SL over FR2, and its analysis is left for future work.} SL  uses the orthogonal frequency division multiplexing (OFDM) waveform with cyclic prefix (CP). In the frequency domain, 12 contiguous OFDM subcarriers form a physical resource block (PRB). The minimum resource allocation unit in frequency is known as a \textit{subchannel} which can comprise $10$ to $100$ PRBs.
In the time domain, a single SL radio frame is of duration $10$ ms and is further divided into $10$ subframes ($1\text{ ms}$ each). Each subframe has multiple time \textit{slots}, the minimum unit of resource allocation in time. SL supports NR's scalable numerology, with slot durations depending on the OFDM sub-carrier spacing (SCS). For FR1 SL, SCS values of \{$15, 30, 60$\} kHz are allowed, corresponding to slot durations of \{$1, 0.5, 0.25$\} ms, respectively. One time slot contains 14 OFDM symbols. One SL transmission instance, known as a transport block (TB), can comprise multiple contiguous subchannels in the same slot. 

\begin{figure}
    \centering    \includegraphics[width=0.75\columnwidth]{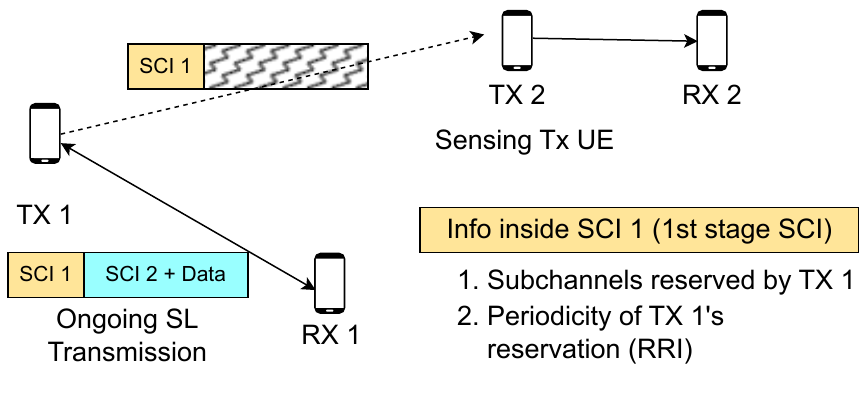}
    \caption{SCI and data transmissions in NR SL}
    \label{fig:SCI_concept}
    \vspace{-0.5cm}
\end{figure}

A SL TB comprises control information followed by data. Sidelink control information (SCI) is sent in two parts: first stage and second stage SCI ($\text{SCI}_1$ and $\text{SCI}_2$). $\text{SCI}_1$ is sent over the physical SL control channel (PSCCH) and can be decoded by any other SL UE if received without error. $\text{SCI}_2$ and the data are sent over the physical SL shared channel (PSSCH) and can only be decoded by the intended receiving UE. We illustrate this in Fig.~\ref{fig:SCI_concept} where TX1-RX1 is an ongoing SL transmission. RX1 can receive and decode the entire TB, while other nearby SL UEs (e.g., TX2) can only decode $\text{SCI}_1$. $\text{SCI}_1$ includes information about the resources reserved by TX1, and its resource reservation interval ($\text{RRI}$), i.e., the transmission periodicity. A reserved resource refers to a set of contiguous subchannels in the frequency domain. If  TX2 receives an SCI from TX1 in slot $x$, indicating that TX1 uses resource ${S}$ with an RRI of $y$, then it knows that TX1 will also occupy the same resource in slots $x + y$, $x + 2y$ and so on. 

\subsection{SL resource allocation procedures at the MAC layer}
\colblue{\textbf{Mode 1}}: This is the centralized resource allocation mode of SL where the NR gNB (base station) controls the scheduling and allocates resources to the SL UEs for their D2D transmissions. Mode 1 requires the SL Tx UE to be in coverage of an NR gNB. To support Mode 1, the AR glasses may rely on the companion if they cannot directly connect to the gNB (possibly due to hardware/battery constraints).

\colblue{\textbf{Mode 2}}: In Mode 2, each SL Tx UE performs a distributed sensing-based procedure to select resources for its transmissions. We consider the semi-persistent scheme (SPS) of Mode 2, whereby the SL Tx UE uses a newly reserved resource for its next $\textit{RRC}$\footnote{$\textit{RRC}$ stands for the resource re-selection counter.} consecutive transmissions in periodic time slots. 

When selecting a new resource, the SL Tx UE defines a time window (\textbf{selection window)} that starts soon after the latest packet arrival time and lasts up to the packet delay budget (PDB). Step $1$ of Mode 2 is the exclusion of resources already reserved by other SL Tx UEs based on the sensing information ($\text{SCI}$s) collected in a recent history of time known as \textbf{sensing window}.  A $\text{SCI}_1$ from another UE is considered for resource exclusion only if its reference signal received power ($\text{RSRP}$) at the sensing UE is above a certain pre-configured threshold. 

Step $2$ involves randomly selecting a suitable resource among the available candidate resources after Step 1. Once a resource is reserved, it can be used by the SL Tx UE over periodic time slots for its next $RRC$ transmissions. Once this reservation expires, i.e., the resource counter goes to $0$, the Tx UE can perform a new resource selection with probability $p_{change}$ or can keep the same resources for another $RRC$ transmissions with probability $1 -p_{change}$.\footnote{For more details on Mode 2, please refer to~\cite{garcia2021tutorial}.} 

Our study includes an analysis and comparison of Mode 1 vs. Mode 2, which is elaborated in Sec.~\ref{genie-assist}.

\section{Methodology: System Model, Assumptions and Simulation Setup}\label{methodology}
In this section, we describe our assumptions, system model, and simulation setup for evaluating the feasibility of using SL for AR. Our goal is to provide directional insights that can serve as a framework for enhancing the technology, despite the possibility that some assumptions may differ in real scenarios.
\subsection{What spectrum should be used for the AR SL?}
There are three potential spectrum options for the glasses-to-companion SL communication: dedicated licensed spectrum (FR1) for XR SL, licensed spectrum (FR1) shared with other cellular traffic, or shared unlicensed spectrum over the 5-7 GHz bands. Each option has its pros and cons, and it is unclear which one the industry will adopt. Here, we make two simplifying assumptions: (1) the AR SL operates in a shared sub-7 GHz spectrum that may be either licensed or unlicensed, with no interference between the glasses-companion SL and the companion-gNB cellular link, as they operate on different bands, and (2) the companion device has two radios, one for the 5G NR Uu link to the gNB and one for the NR SL.
\subsection{Modeling external interference}\label{external}
AR SL operating on either licensed or unlicensed spectrum will experience interference from other cellular or WiFi/NR-U users. To consider such interference, we assume that external traffic occupies $X\%$ of the available airtime, i.e., for each time slot, there is an independent probability of $X/100$ that it will be occupied by an external user, making it unavailable for AR users to use for SL transmission. If an AR user had already scheduled a transmission during such an occupied slot, they must postpone it to the next reserved slot. The underlying assumption here is that AR users can sense the channel for external interference to detect slots that are already occupied. \colblue{SL devices already use channel sensing to detect ongoing SL transmissions in the network and decode their SCIs. To detect any external interference occupying the channel, a simple energy-based detection threshold similar to the one used in Wi-Fi systems can be employed.} 
 \subsection{Deployment Scenario}\label{deployment}
\colr{We design a deployment scenario that can represent several indoor AR use cases, such as an office space, a single floor of a multi-family apartment complex, an AR gaming arcade, or an exhibition/museum space with visitors wearing AR glasses. We model AR users as a pair of SL UEs (companion device and AR glasses) deployed in a  \SI{20}{\metre}$\times$\SI{20}{\metre} 2D square grid with random locations and orientations. We vary the number of users inside the grid from $1$ to $20$ to cover scenarios with light ($1$ to $5$), moderate ($5$ to $10$), or heavy ($5$ to $20$) user density. The glasses-to-companion distance is uniformly distributed between \SI{1}{\metre} to \SI{2}{\metre} for short-range D2D communication.}

\subsection{System Model}
We use the 3GPP indoor office channel model for our study. The path loss is modeled by the following equation:
\begin{align*}
    PL = 32.4 + 17.3\log_{10}{d} + 20\log_{10}{f_c},
\end{align*}
where, $d$ is the 2D Euclidean distance between the transmitter and the receiver, and $f_c$ is the carrier frequency ($6$ GHz). We assume a constant transmit power ($P_{tx}=14$ dBm) and that omnidirectional beams are used for transmission/reception. The received signal power is modeled as:
\begin{align*}
    P_{rx} = G_{rx}*RL_{rx}*PL*G_{tx}*RL_{tx}*P_{tx}
\end{align*}
where, $G_{tx}$ and $G_{rx}$ are the transmit and receive antenna gains. $RL_{tx}$ and $RL_{rx}$ are the transmit and receive RF losses. $PL$ is the pathloss as explained above. We neglect shadowing and body blockage effects, with a worse performance expected, if included. We aim to identify key SL design bottlenecks rather than perform rigorous quantitative analysis, which requires more precise channel modeling and is left for future study.
We consider a network with $N$ SL Tx-Rx pairs, i.e., the AR users. Based on each user's MAC layer scheduling, only a subset of the $N$ users perform SL transmission in a given time slot and frequency subchannel (active users). Let the number of such ``active'' SL pairs be $N_{active}$. Let $P_{rx}^{i,j}$ be the received signal power at $RX_{i}$ from a transmission initiated at $Tx_{j}$, where $P_{rx}$ is as defined above. Let $P_{\text{noise}}$ be the noise power, then the signal-to-interference plus noise ratio (SINR) of the transmission of the $i-th$ Tx-Rx pair is calculated as:
\begin{align*}
    SINR = \frac{P_{rx}^{i,i}}{\sum_{j\neq i, i=1}^{i=N_{active}}{P_{rx}^{i,j}} + P_{\text{noise}}}
\end{align*}
\subsection{Traffic model}
\label{traffic}
3GPP recommends a quasi-periodic traffic generation model for XR apps, where a burst of traffic is generated at the beginning of each period, or ``traffic interval'', with a rate of 1/T Hz~\cite{petrov2022extended, 3gpp.38.838}. Our simulations use a default rate of 45 Hz, which can range from 30-120 Hz. The burst includes both glasses-to-companion and companion-to-glasses traffic, with the amount depending on the use case, e.g., a $20$ Mbps requirement results in $55$kB of traffic every $22$ ms. Traffic generated at the start of an interval must be transferred within the same period, imposing a latency limit on every SL transmission. A packet is dropped if it cannot be scheduled within this limit. Promptly completing the SL transfer allows the AR glasses to enter sleep mode and save power, as discussed in~\cite{kim2021ue}.
\subsection{Bi-directional traffic support}
In the unicast operation of SL Mode 2, the Tx UE performs resource selection for its transmissions to its corresponding Rx UE. However, in AR, the SL traffic is bidirectional, where both the AR glasses and the companion can act as the Tx or Rx UE. We consider two options: (1) the companion device performs sensing and resource allocation and informs the AR glasses through SCI, and (2) independent sensing and resource allocation by both devices. We select Option 1, where the AR glasses do not have to perform sensing and resource selection, saving energy. Coordination between the companion device and the glasses for resource allocation should be feasible in practice. Option 2 will result in performance degradation due to collisions as both devices compete for channel access. 

\subsection{SL resource allocation: Mode 1 vs. Mode 2}\label{genie-assist}
\colblue{In Mode 1 (centralized), the gNB controls the scheduling and has more information about the network than the SL UEs performing sensing-based resource selection in the case of Mode 2. A comparative analysis of both modes is desirable. Before describing our simulation approach, we first list the following key features of Mode 1:
\begin{itemize}
    \item The gNB uses its knowledge of future resource allocations for all SL users to prevent collisions among them. 
    \item In Mode 2, the SL Tx UE cannot sense the channel in a time slot in which it is also transmitting data unless the device can support full duplex operation.  Mode 1 \textbf{does not} have this constraint, as the gNB performs the resource allocation. The gNB has all the information on the resources allocated to SL users.
    \item The gNB can utilize the location information of all SL UEs to efficiently reuse resources among spatially separated SL Tx-Rx pairs.
\end{itemize}
In our simulation, we model Mode 1 by augmenting our Mode 2 implementation with a ``genie'' assisted approach. The genie provides all SL Mode 2 UEs with additional information about the network, which is used to emulate the aforementioned features unique to Mode 1. This allows us to approximate Mode 1 performance. More details of this approach are provided in~\cite{srivastava2023enhanced}, where it was used to get an upper bound for SL Mode 2 performance. We compare Mode 1 and Mode 2 performance results in Section~\ref{results}.}
\subsection{MCS configuration and Link adaptation}
We use three representational MCS values (Table~\ref{tab:mcs_config}) from Table $5.1.3.1$-$1$ of 3GPP standard 38.214~\cite{3gpp.38.214}, with MCS 19 for high data rates and MCS 4 for reliable data transfer in poor link conditions. The companion estimates interference using the SCIs it received during sensing. It calculates the expected SINR for its SL transmission over the selected resources and maps the SINR to a block error rate (BLER). MCS is then adapted to meet a desired BLER target ($1\%$ by default\footnote{The BLER target is device vendor implementation dependent. A higher value may be configured if it can be tolerated by the target AR use cases.}). 

\colblue{For the SINR to BLER mapping, we use the physical layer (PHY) abstraction model from~\cite{lagen2020new}, which captures the 5G NR specifications related to low-density parity check (LDPC) channel coding, such as LDPC base graph selection and code block segmentation. It uses link-level simulations and the exponential effective SINR mapping (EESM) method to obtain the SINR vs. BLER data. For more details, you may refer to the original work in~\cite{lagen2020new}.}

\begin{table}[h]
\vspace{-1cm}
\begin{tabular}{ |p{0.8cm}||p{1.2cm}||p{1.5cm}||p{1.1cm}||p{2cm}| }
 \hline
 MCS Index& Modulation Order& Code Rate ($R\times1024$)& Spectral efficiency& ~{Data rate at $100$ MHz BW (Mbps)} \\
 \hline
 4& 2& 308& 0.6016& 60.16 \\
 \hline
 11& 4& 378& 1.4766& 147.66\\
 \hline
 19& 6& 517& 3.0293& 302.93\\
 \hline
\end{tabular}
\caption{\protect\label{tab:mcs_config}MCS configurations}
\end{table}
\subsection{SL Simulation parameters}
Our simulation parameters comply with 3GPP specifications and are shown in Table~\ref{tab:sim_params}\footnote{Our slot time is $0.5$ ms, and SL UEs follow the half-duplex sensing constraint~\cite{garcia2021tutorial}. Under these assumptions, an RRI of $1$ ms would restrict the UE's choice of resources, making the use of sensing-based Mode 2 meaningless and possibly causing high packet errors. Thus, we select $RRI = 2$ ms, which is the next minimum permissible value, to minimize SL transfer latency.\label{footnote_4}}. For details on SL Mode 2 parameters, please refer to~\cite{garcia2021tutorial}. The reservation size equals the number of frequency sub-channels, optimizing the AR D2D transfer latency by using the entire bandwidth in a single slot. 
\begin{table}[h]
\begin{tabular}{ |p{5.2cm}||p{2.4cm}| }
 \hline
 \textbf{Simulation Parameter}& \textbf{Value} \\
 \hline
 Available bandwidth (MHz)& $20/40/60/80/100$\\
 \hline
 Sub-carrier spacing (SCS)& $30$ kHz \\
 \hline
 Slot Time& $0.5$ ms\\
 \hline
 Subchannel size & $15$ PRBs\\
 \hline
 Resource Reservation Interval (RRI) & $2$ ms\\
 \hline
 Resource re-selection counter& $\stackrel{}{\sim}U[25,75]$\\
 \hline
 Probability of changing resources in Mode 2& $0.5$\\
 \hline
 RSRP threshold in Mode 2 & $-96$ dBm\\
 \hline
 Mode 2 sensing window & $100$ ms\\
 \hline
\end{tabular}
\caption{\protect\label{tab:sim_params}SL simulation parameters}
\vspace{-0.25cm}
\end{table}

\subsection{Performance evaluation metrics}
\noindent
\textbf{Packet reception ratio (PRR)}: Each SL packet (TB) can have three possible outcomes: (1) \textit{success}: Successful reception at the Rx~\footnote{Our simulator maps SINR values to BLER~\cite{lagen2020new} and determines the success of a TB's reception by flipping a biased coin with the BLER as the probability.}, (2) \textit{$Rx_{\text{failure}}$}: Packet reception error at the Rx side, and (3) \textit{$\text{drop}$}: Packet is dropped at the Tx side due to exceeding its deadline, i.e., $22$ ms latency limit. A user's PRR is the ratio of its \textit{$\text{success}$} packets to its total number of transmissions. \\
\textbf{Latency} for a SL packet is $T_{\text{reception}} - T_{\text{arrival}}$, the time difference between its reception at the Rx and generation at the Tx. \colb{AR traffic is generated every $22$ ms, and the total ``on-time'' of the glasses-companion link is determined by the latency of the last SL transmission of an AR user in each  traffic interval. This is the key metric since it determines whether the SL connection can satisfy the AR glasses' thermal constraints and whether the bursty AR traffic can meet its total latency requirements. In our results, we refer to this metric simply as ``Latency''.} 

\subsection{How do we judge AR use case satisfaction ?}
Without limiting ourselves to specific product specifications, we use a semi-quantitative approach to evaluate the feasibility of SL for AR. We use two criteria - end-to-end latency impact and thermal limit of the AR glasses. AR involves multiple processing steps~\footnote{This includes data capture using sensors, computation-heavy tasks like simultaneous localization and mapping (SLAM), hand tracking and rendering, the wireless D2D transfer, and the final display of the rendered object(s).}, and each step incurs latency, including the bidirectional glasses-to-companion SL transfer. Previous studies indicate that wireless transfer latency requirements for AR range from 5-50 ms, depending on the specific use case~\cite{alriksson2021xr}. We classify use cases into three broad categories:\\
\textbf{Class A}- Interaction-based use cases with world-locked rendering (WLR): The user interacts with the AR content in real time. WLR is supported, where content is placed at a fixed position in the environment, for comfortable user experience~\cite{HLR,zhao2023evaluating}. \colblue{Some examples of these use cases include holographic AR board games, online AR shopping, and AR-guided remote assistance for industry services~\cite{3gpp.26.928}.} WLR requires advanced processing, resulting in a stricter latency budget. \colb{We set} the round-trip D2D wireless latency target for Class A to be around $15$-$20$ ms. \\
\textbf{Class B} - Interaction-based use cases without WLR: The latency requirements are around $30-40$ ms. \colblue{An example use case for this category is 3D AR video calling.}\\
\textbf{Class C} - Simple preview, file transfer, or messaging: Less latency-sensitive use cases, with $100$ms - $1$sec. requirements. \colblue{Examples include previewing 3D models of new products, 3D image/video sharing, and instant messaging using AR prompts or virtual keyboard~\cite{3gpp.26.928, 9288380}.}

SL must operate within the thermal constraints of the AR glasses, which limits the duty cycle of the wireless radio. Specifically, if the thermal limit is $X\%$, all SL transfers must be completed within $(X/100)*T$ ms of a periodic traffic interval of $T$ ms. The radio will then enter sleep mode for the remaining time to save power. We set the thermal limit to 50\% as a rough anchor point, recognizing that its accuracy may depend on product specifications and design choices.


\section{Results}\label{results}
For each result, we ran $20$ simulations, each with a different random distribution of UEs for the deployment scenario described in Sec.~\ref{deployment}. Each simulation run lasted for $10$ seconds. The results represent the system-wide average PRR and latency for all AR users and across the 20 distributions.
\subsection{Baseline performance @ 100 MHz BW, no BG interference}\label{baseline}
\begin{figure}[t]
\centering
\begin{subfigure}{0.49\linewidth}
    \centering
    \includegraphics[width = \textwidth]{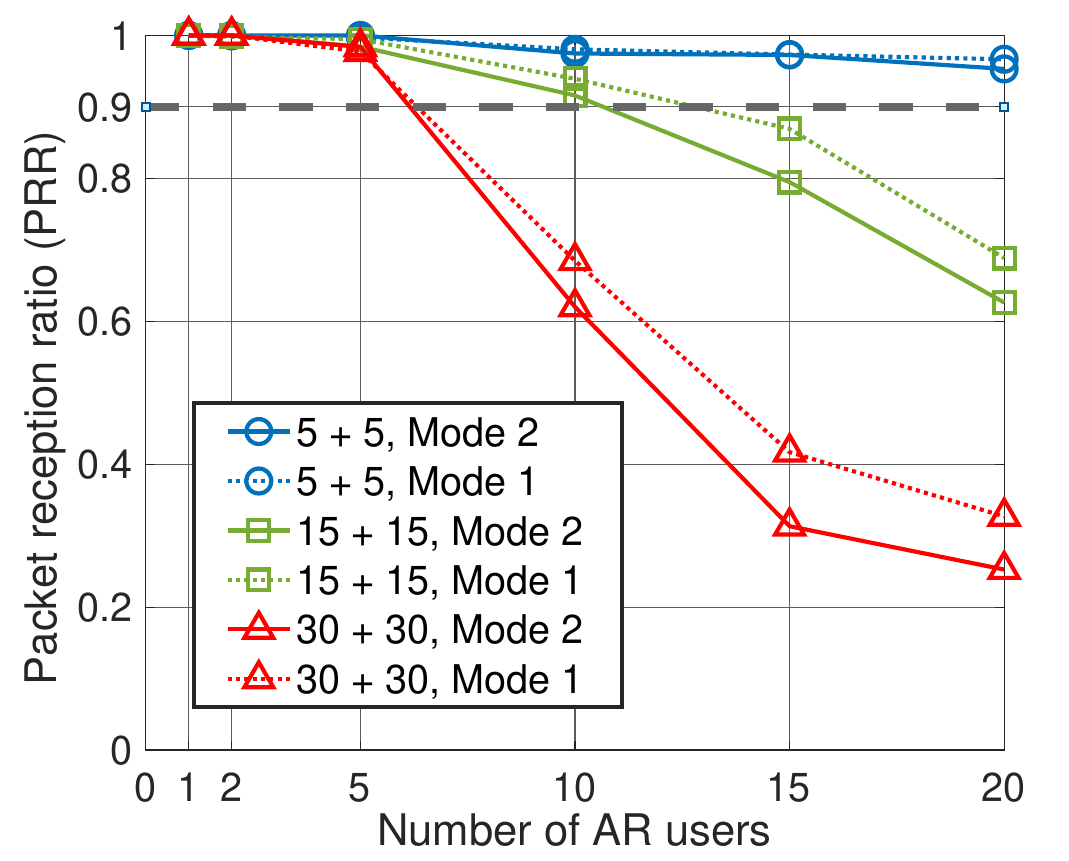}
    \caption{PRR}
    \label{fig:PRR_1}
\end{subfigure}
\begin{subfigure}{0.49\linewidth}
    \centering
    \includegraphics[width=\textwidth]{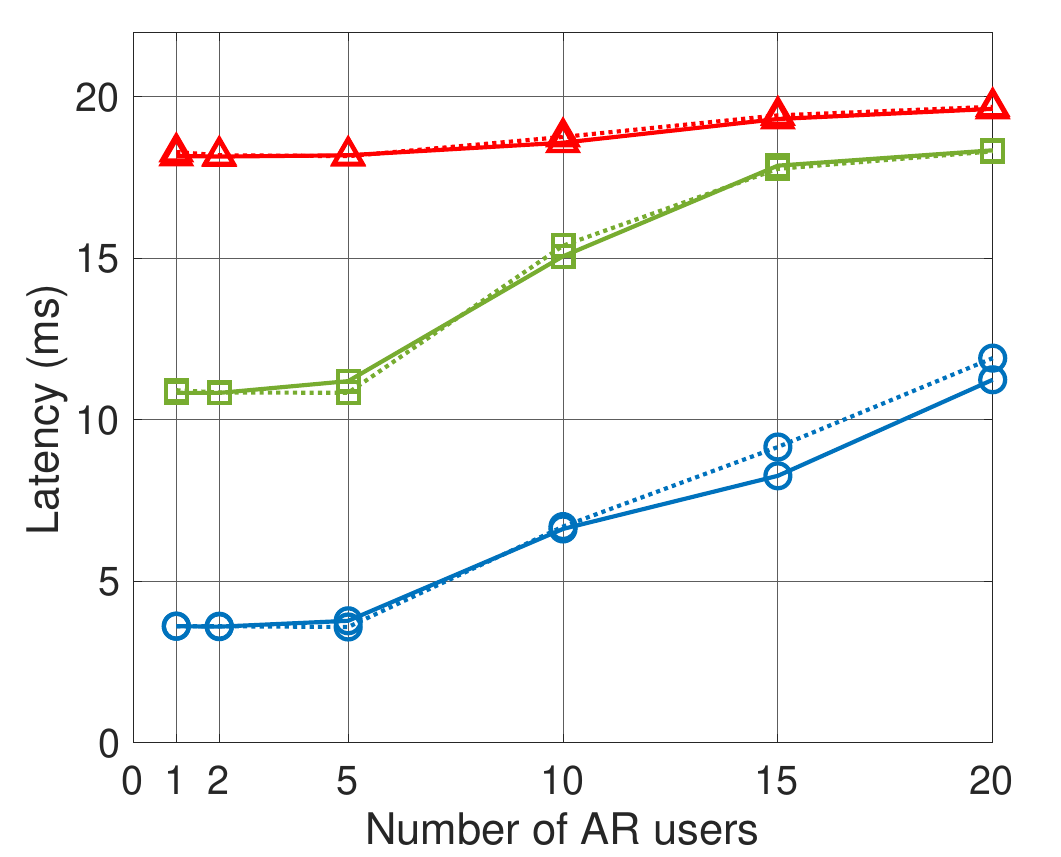}
    \caption{Latency}
    \label{fig:latency_1}
\end{subfigure}
\caption{Performance of the AR glasses' SL connection at varying traffic loads - $100$ MHz BW, no external interference}
\label{fig:result_1}
\vspace{-0.6cm}
\end{figure}

\textbf{Fig.~\ref{fig:result_1}} shows the performance of the AR glasses' SL connection in a multi-AR user scenario. The available bandwidth is $100$ MHz, with no background interference. We simulate SL Mode 2 and our genie-assisted approximation of Mode 1 from Sec.~\ref{genie-assist}. With a traffic load of $5+5$ Mbps\footnote{$5+5$ denotes glasses-to-companion (G2C) and companion-to-glasses traffic. Our results with symmetric traffic still provide a representative view of performance with asymmetric traffic and the same total load. As the G2C link is usually less loaded, our choice can represent the worst-case scenario (for the power-constrained AR glasses) under the same total load.}, \colr{SL can support up to 15 users with high reliability and latency below $10$ ms. This meets  all use case requirements and stays within the thermal limit~\footnote{The thermal limit (maximum ON duration) is $50$\% of $T$($22$ms) = $11$ms}. However, as the traffic load increases, the PRR goes down, and latency increases, degrading performance and exceeding the thermal limit.} With $15+15$ Mbps traffic, the requirements of class A use cases are not met, and only class B and C use cases can be supported for less than $10$ users. With even more traffic ($30+30$), only class C use cases can be supported by allowing SL TBs to be queued for transmission in subsequent intervals. We also see that Mode 1 and Mode 2 have similar latency, while Mode 1 shows better reliability under heavy traffic with many users~\footnote{\colblue{Our results show the user plane latency but do not include the control plane latency overhead needed for exchanging resource allocation information between the gNB and the SL UE in Mode 1. This is left for future work.}}. Despite this, the trends for both schemes are similar, indicating that our results for Mode 2 can be a good indicator of Mode 1 performance. \colblue{Hence, the results presented in subsequent sections will only consider Mode 2 unless specified otherwise.}

\subsection{Impact of available bandwidth and external interference}
\begin{figure*}
\centering
\begin{subfigure}{0.21\linewidth}
    \includegraphics[width = \textwidth]{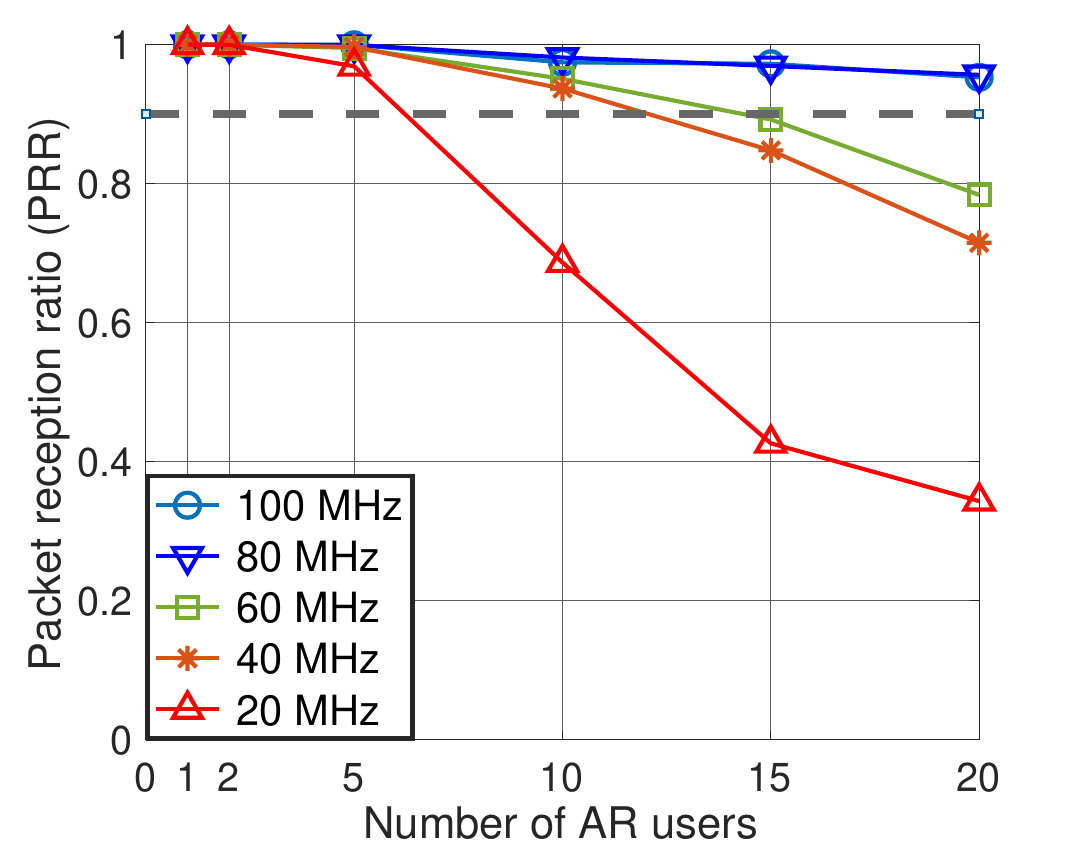}
    \caption{PRR - 5+5 Mbps}
    \label{fig:PRR_5_BW}
\end{subfigure}
\begin{subfigure}{0.21\linewidth}
\hspace*{-0.5cm}
    \includegraphics[width = \textwidth]{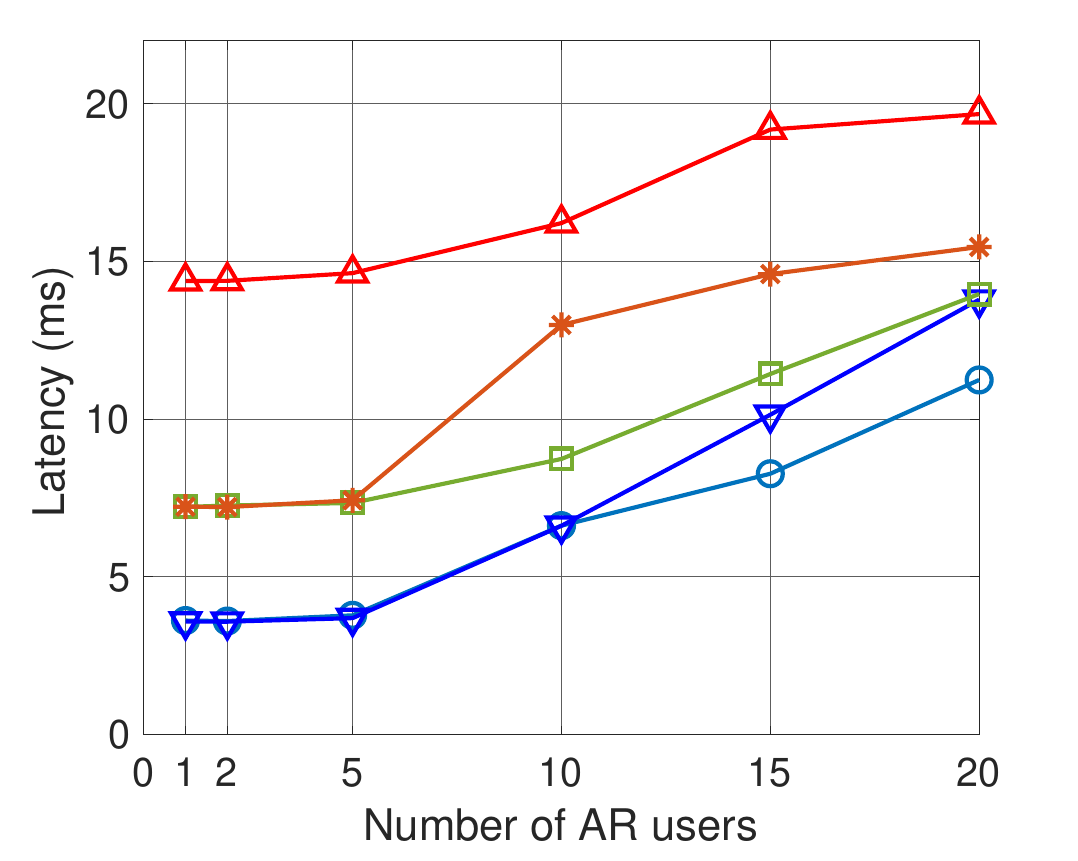}
    \caption{Latency - 5+5 Mbps}
    \label{fig:latency_5_BW}
\end{subfigure}
\begin{subfigure}{0.21\linewidth}

    \includegraphics[width = \textwidth]{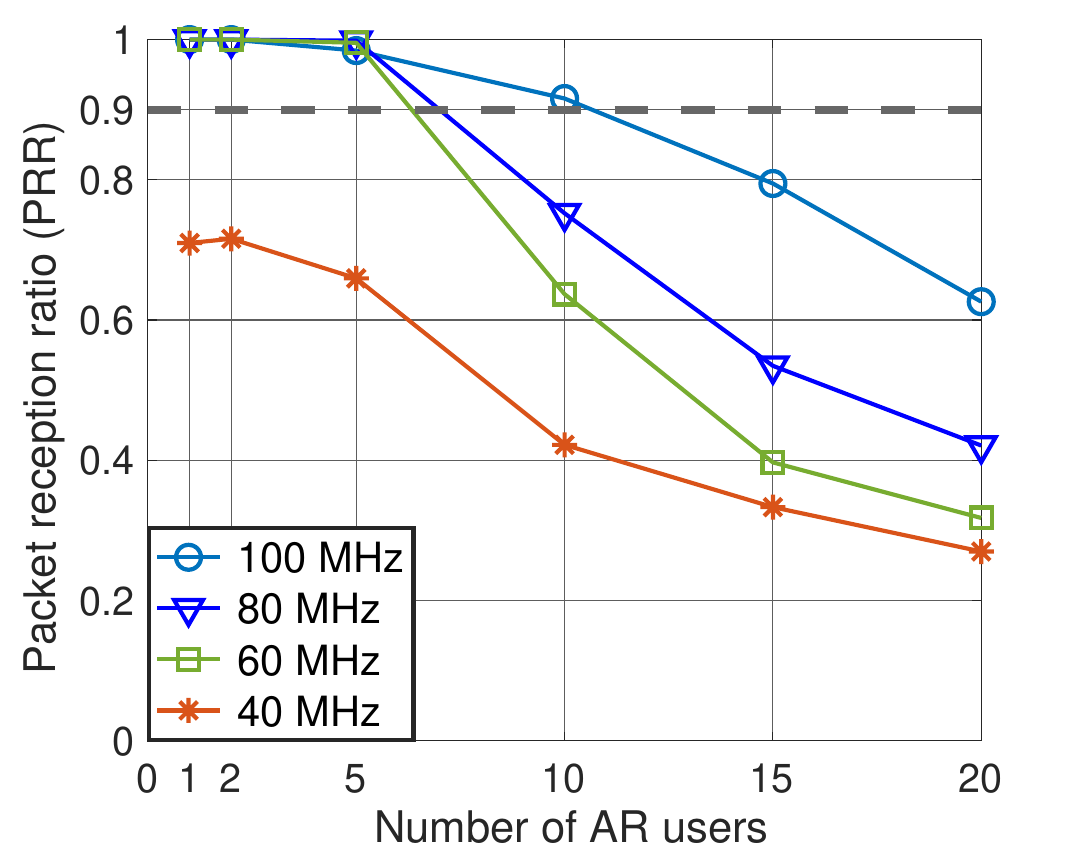}
    \caption{PRR - 15+15 Mbps}
    \label{fig:PRR_15_BW}
\end{subfigure}
\begin{subfigure}{0.21\linewidth}
    \includegraphics[width = \textwidth]{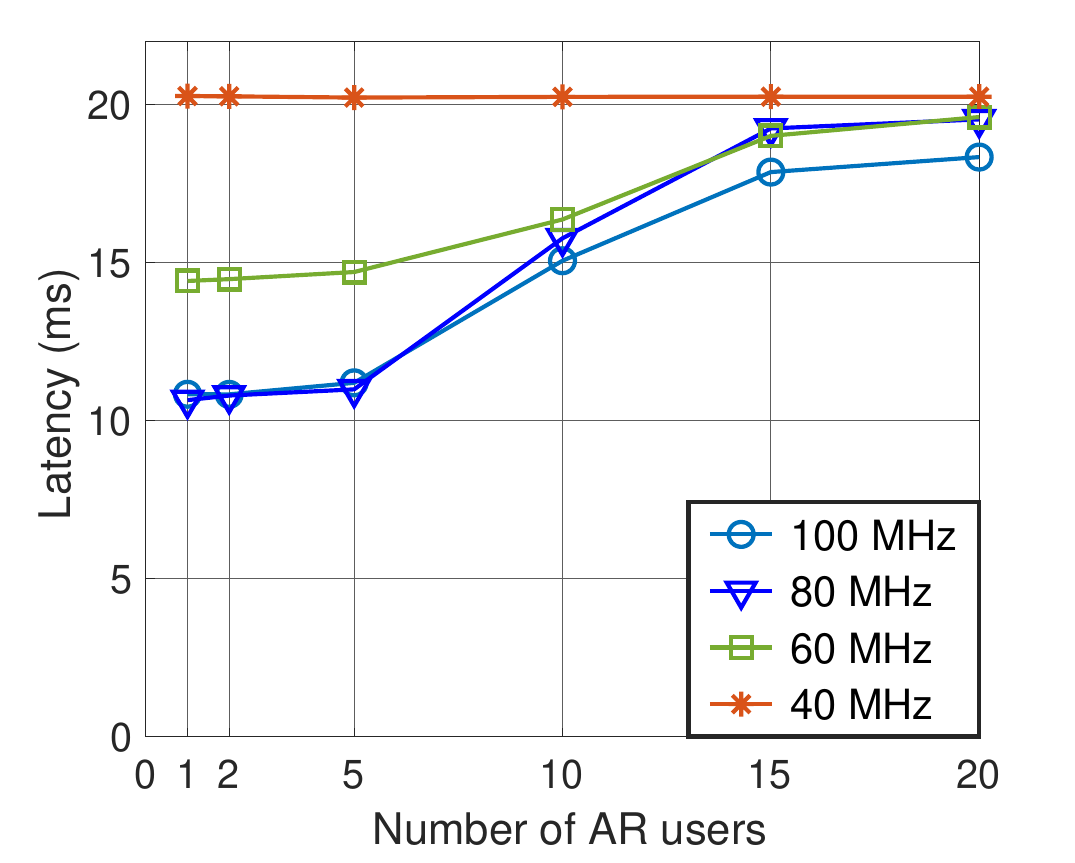}
    \caption{Latency - 15+15 Mbps}
    \label{fig:latency_15_BW}
\end{subfigure}
\caption{Impact of available bandwidth on SL performance ($X = 0$ \%)}
\label{fig:result_BW}
\vspace{-0.25cm}
\end{figure*}

\begin{figure*}
\centering
\begin{subfigure}{0.21\linewidth}
    \includegraphics[width=\textwidth]{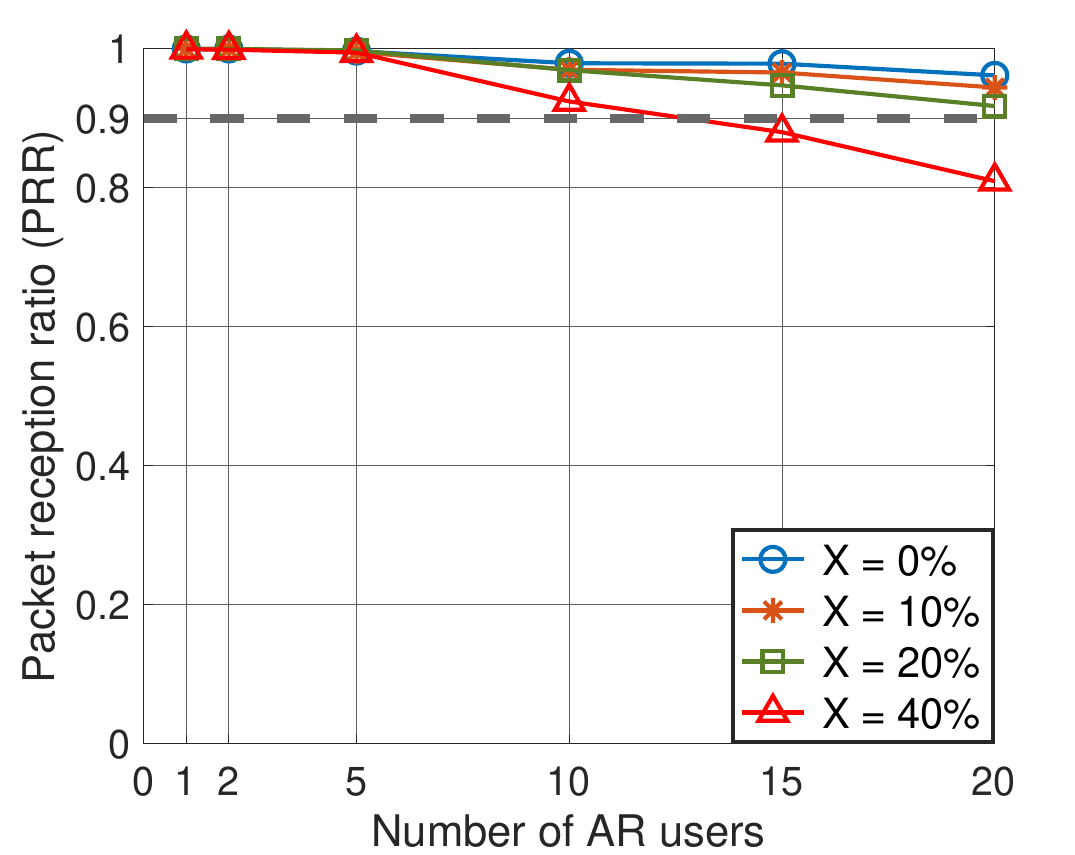}
    \caption{PRR - 5+5 Mbps}
    \label{fig:PRR_5_if}
\end{subfigure}
\begin{subfigure}{0.21\linewidth}
    \includegraphics[width=\textwidth]{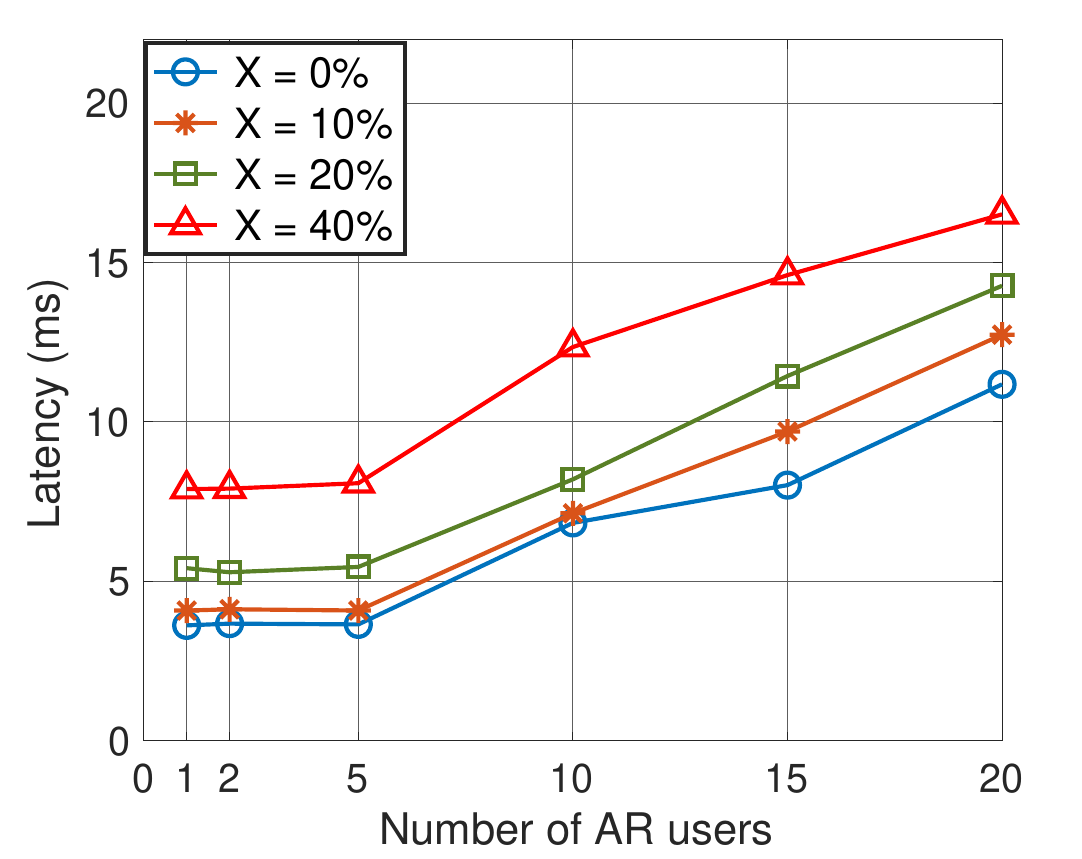}
    \caption{Latency - 5+5 Mbps}
    \label{fig:latency_5_if}
\end{subfigure}
\begin{subfigure}{0.21\linewidth}
    \includegraphics[width=\textwidth]{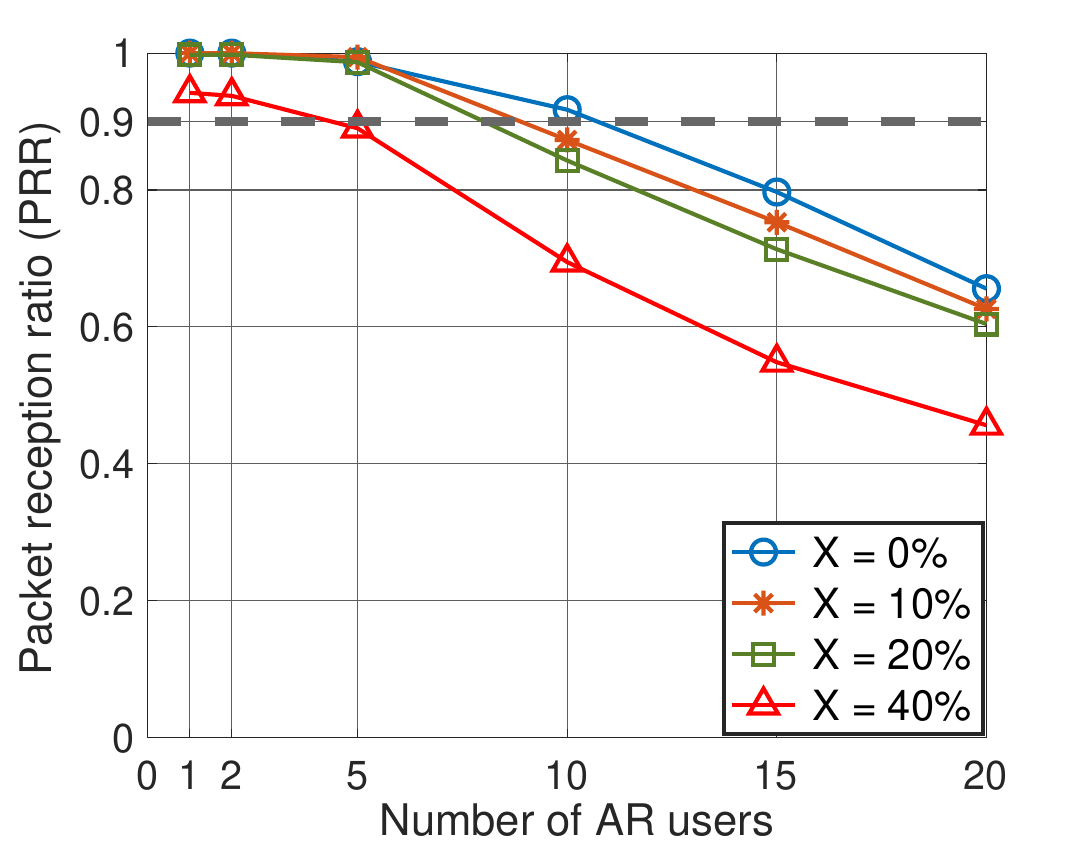}
    \caption{PRR - 15+15 Mbps}
    \label{fig:PRR_15_if}
\end{subfigure}
\begin{subfigure}{0.21\linewidth}
    \includegraphics[width=\textwidth]{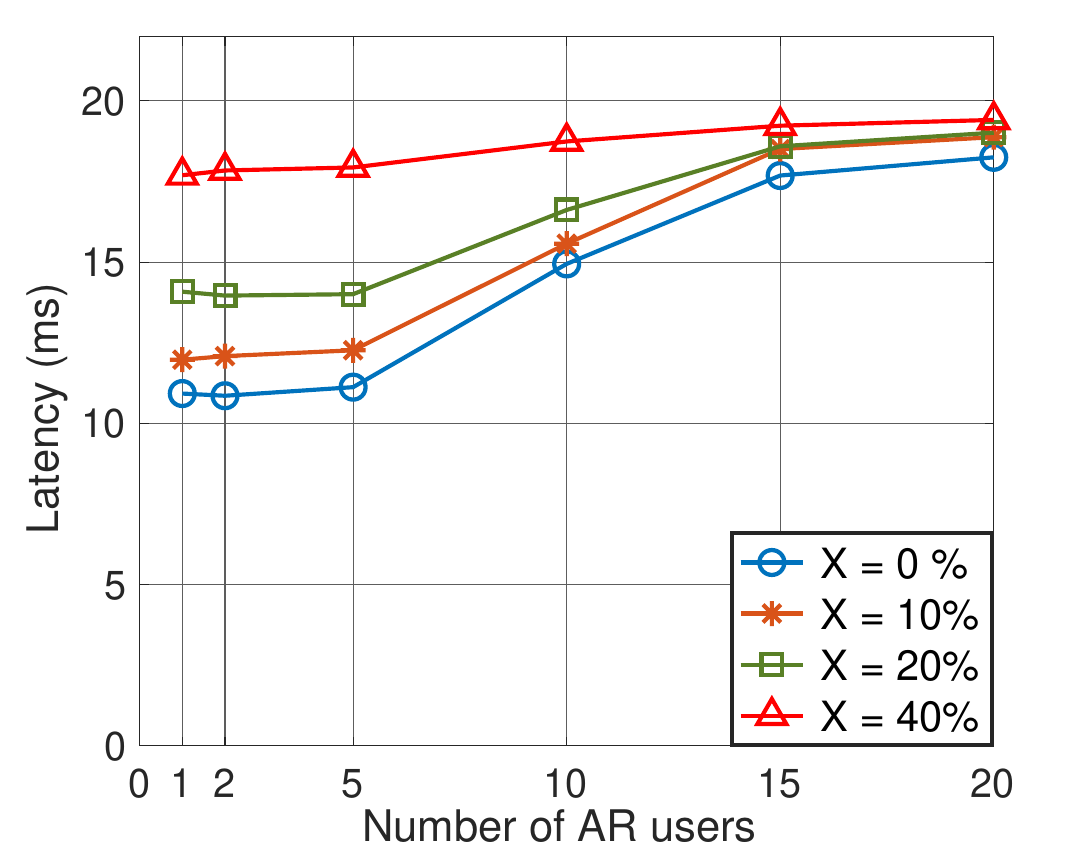}
    \caption{Latency - 15+15 Mbps}
    \label{fig:latency_15_if}
\end{subfigure}
\caption{Impact of external interference on SL performance ($BW = 100$ MHz)}
\label{fig:result_IF}
\vspace{-0.25cm}
\end{figure*}

In \textbf{Fig.~\ref{fig:result_BW}}, we look at how the amount of bandwidth (BW) available for SL impacts performance. \colr{A lower BW implies more congestion in the network, leading to lower PRRs and higher latencies.} At a traffic load of  $5+5$ Mbps, at least 60 MHz of BW is required to support all use cases for up to 10 users (Fig.~\ref{fig:PRR_5_BW} and Fig.~\ref{fig:latency_5_BW}). Up to 5 users can be supported with 40 MHz BW, and 20 MHz is insufficient for any use case due to the thermal limit. For $15+15$ Mbps, we need a minimum BW of $80$ MHz to support all use cases for up to 5 users. So far, no external interference has been considered. In \textbf{Fig.~\ref{fig:result_IF}}, we fix $BW$ to $100\text{ MHz}$ and vary the external interference. For $5+5$ Mbps, up to $10\%$ interference can be tolerated without significant degradation in KPIs, with higher tolerance if there are $\leq 5$ users. For $15+15$ Mbps, only low interference levels ($0-10\%$) are acceptable, while high interference ($40\%$) results in excessive SL latency beyond the thermal limit.

\subsection{Proposed enhancements to SL resource allocation}\label{enhancement}
\begin{figure}[h]
    \centering    \includegraphics[width=\columnwidth]{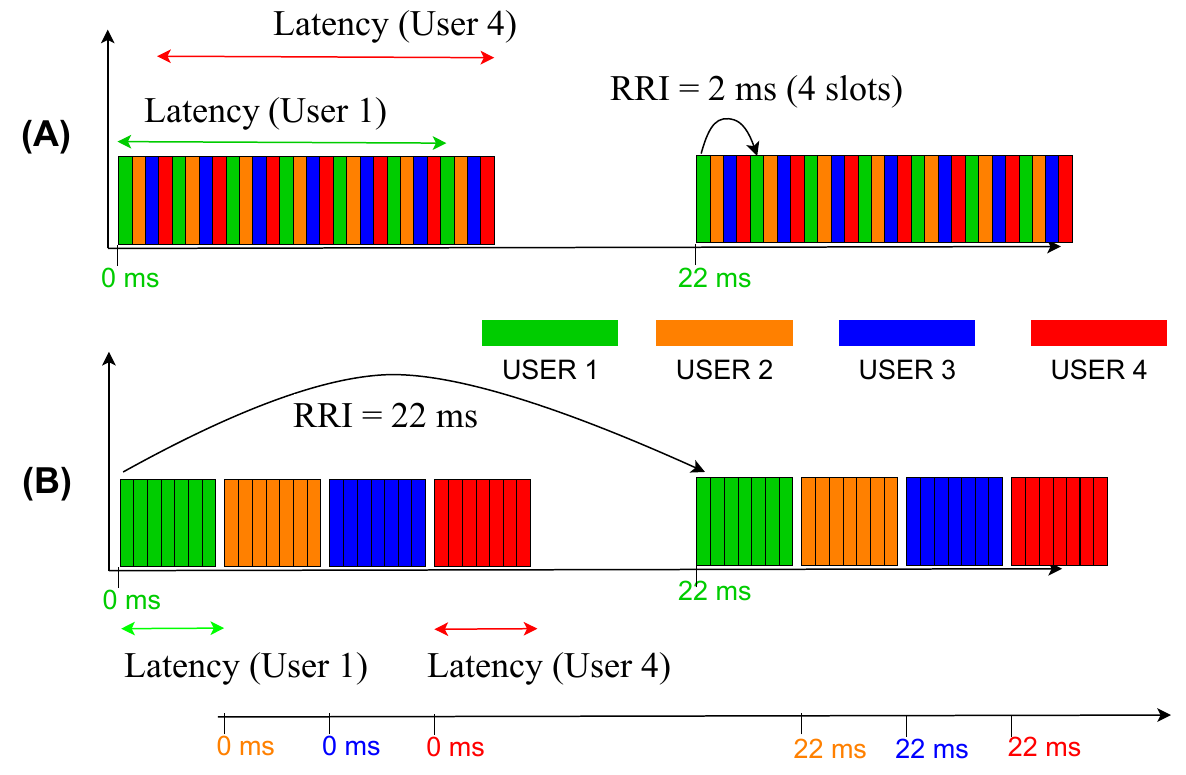}
    \caption{\colr{Using multiple active reservations to reduce SL transfer latency for AR users, (A): Legacy SPS, (B): MAR scheme}}
    \label{fig:MAR_concept}
    \vspace{-0.25cm}
\end{figure}

\begin{figure}[h]
\begin{subfigure}{0.49\linewidth}
    \includegraphics[width=\textwidth]{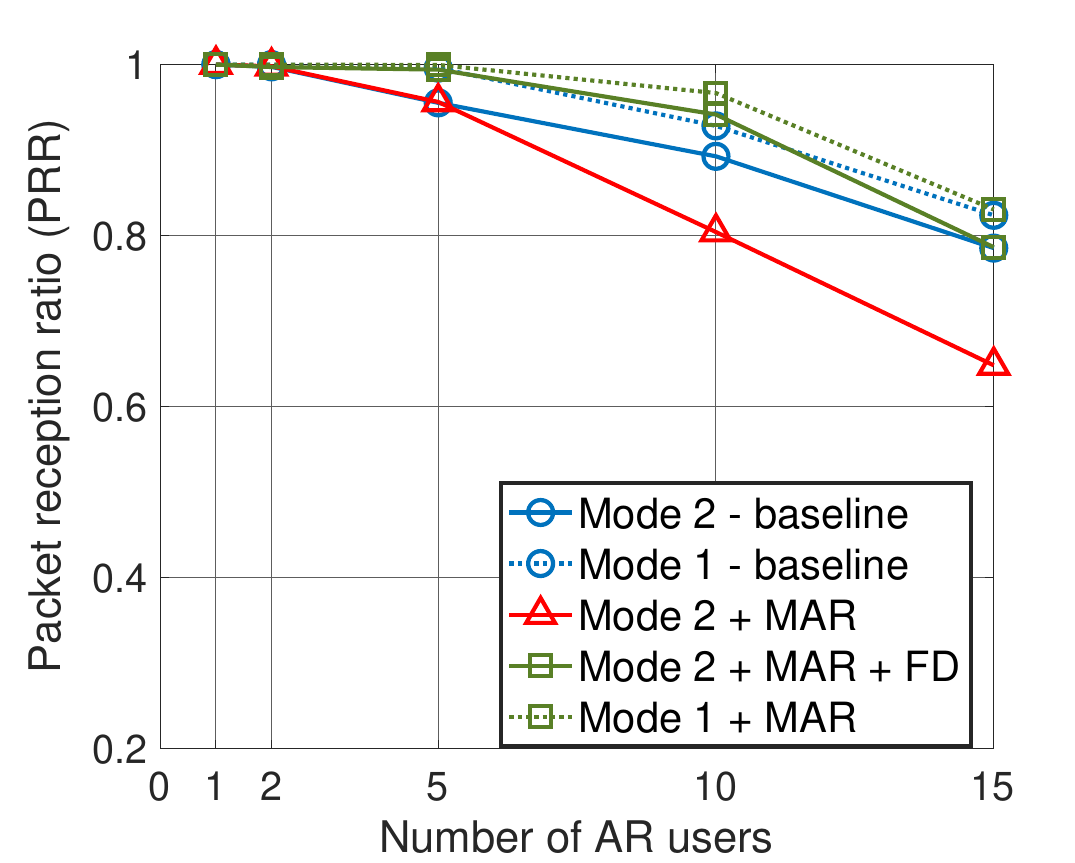}
    \caption{PRR}
    \label{fig:first}
\end{subfigure}
\hfill
\begin{subfigure}{0.49\linewidth}
    \includegraphics[width=\textwidth]{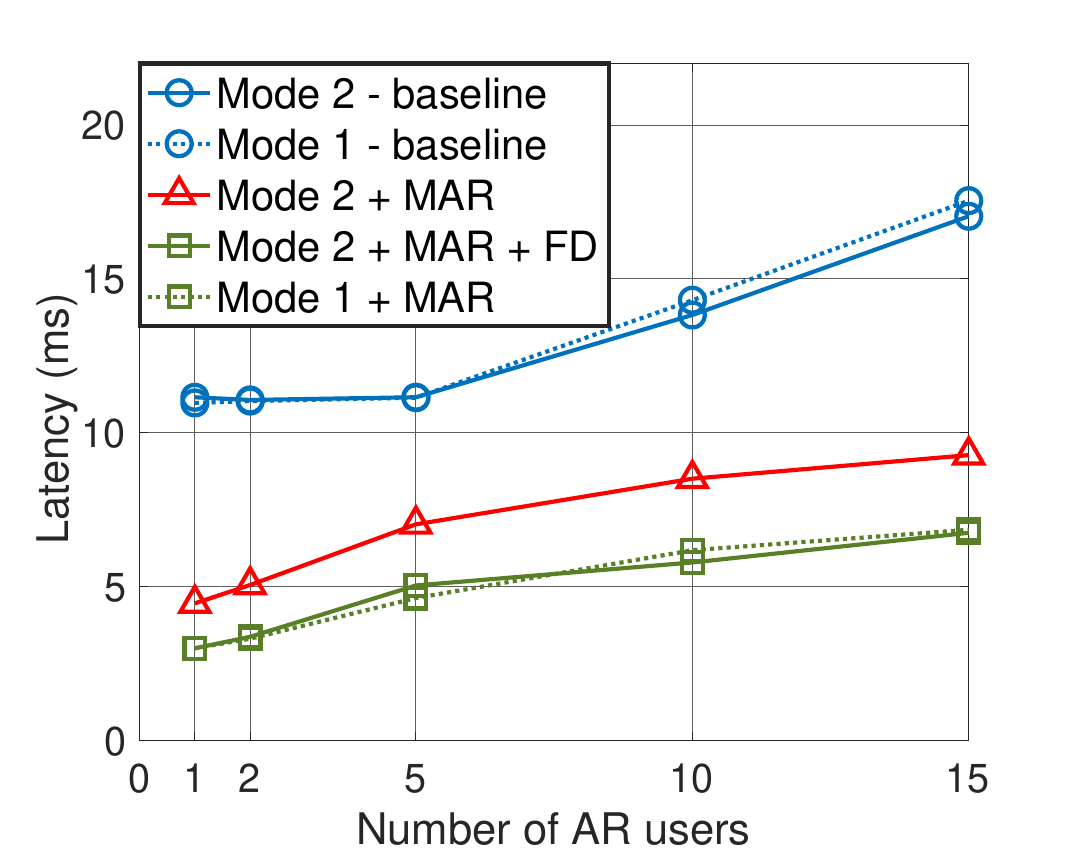}
    \caption{Latency}
    \label{fig:second}
\end{subfigure}
\hfill
\caption{Performance of the proposed enhancements - Multiple active reservations (\textbf{MAR}) and Full duplex sensing (\textbf{FD})}
\label{fig:SL_enhancements}
\vspace{-0.25cm}
\end{figure}

In the results so far, we observe high SL transfer latencies in many scenarios, particularly the ones with high traffic loads, limited bandwidth, and external interference. \colb{The main bottleneck here is the periodic nature of resource scheduling in the semi-persistent scheme (SPS) of SL Mode 2, which is not well suited to the bursty nature of AR traffic. With a scheduling periodicity ($RRI$) of $2$ ms, the latency of the last TB sent in a traffic interval, i.e., the overall SL transfer latency for the AR user becomes high.} \colr{This is illustrated in Fig.~\ref{fig:MAR_concept} (A).} 3GPP has also defined a ``dynamic scheme'', wherein the SL Tx UE selects a new \colb{time-frequency resource} for each TB. This is more suited for aperiodic traffic. However, if all SL devices in a network use the dynamic scheme, the channel access becomes equivalent to random resource selection, degrading reliability. As discussed in Note~\ref{footnote_4}, an RRI lower than $2$ ms is not feasible for our system. We propose two enhancements to SL Mode 2 that retain the benefits of sensing-based channel access while being more suitable for AR traffic:\\
(1) Use of \textbf{multiple active SL reservations (MAR)}: Instead of selecting a single periodic resource with $RRI=2$ ms (periodicity of $4$ time slots) \colr{as in Fig.~\ref{fig:MAR_concept} (A)}, we now allow a SL Mode 2 scheduling UE (the companion device) to select multiple resources in consecutive time slots. \colr{This allows an individual AR user's D2D transfer to complete quickly, reducing latency. In addition, the SL radio on the glasses will be active for a much shorter duration and can go to sleep after the transfer completes, thus alleviating the thermal impact issue.} The MAR solution is depicted in \colr{Fig.~\ref{fig:MAR_concept} (B)}, where users can now use consecutive time slots to complete their SL transmissions. The $RRI$ is set to $22$ ms, corresponding to the traffic interval duration. Thus, a single UE can now select a ``batch'' of resources in consecutive time slots and use the same batch in longer periodic intervals to meet its traffic requirements. We simulate this scheme and observe its performance in \textbf{Fig.~\ref{fig:SL_enhancements}} for $15+15$ Mbps traffic. Our \textbf{baselines} for comparison are the conventional SL Mode 1 and Mode 2 schemes with periodic reservations. We see that SL Mode 2 with MAR has a significantly reduced latency; however, the PRR of Mode 2 $+$ MAR is much lower than the baseline Mode 2 scheme~\footnote{\colblue{Persistent collisions can occur in the SPS scheme of Mode 2~\cite{todisco2021performance}. This gets exacerbated with MAR, where users can hold on to a batch of reservations for a longer period. Mode 2 configurations, such as the $RRC$ and $p_{\text{change}}$, can be optimized to better suit MAR. However, this is left for future work.}}. With Mode 1, we observe that the MAR scheme can offer both low latency and high reliability.\\
(2) \textbf{Full-duplex (FD) sensing}: For Mode 2 SL devices, half-duplex operation prevents sensing during transmission. There is no such constraint in Mode 1 (centralized), which does not rely on sensing. This constraint can be eliminated from Mode 2 if the SL device performing sensing has FD capabilities. In Fig.~\ref{fig:SL_enhancements}, we see that a combination of MAR and FD greatly improves SL Mode 2 reliability and latency performance.

\colb{Implementing MAR should be feasible in practice; however, it remains to be seen whether its benefits can generalize to more use cases, thereby motivating efforts for standardization. The coexistence of MAR users with those using periodic SL reservations is an open question. FD sensing improved SL resource allocation for V2X in~\cite{campolo2019full}, and we demonstrated its benefits to SL for AR. However, designing practical FD radios with self-interference cancellation is an ongoing issue.}

\section{Conclusions}\label{conclude}
We analyze the feasibility of using sidelink (cellular D2D communication) for the wireless connection between a user's AR glasses and a companion device (such as their smartphone) used for computation offloading. We use system-level simulations to model the PHY and MAC layers of 5G NR SL and simulate realistic AR traffic and deployment scenarios. We find that the current SL design struggles to support advanced AR use cases, particularly those requiring real-time interaction, at medium to high traffic loads. The periodic scheduling of V2X SL does not align with AR glasses' latency and power consumption KPIs. To address these issues, we propose two enhancements to SL Mode 2: multiple active reservations (MAR) and full-duplex sensing. MAR enables SL to transmit continuously, reducing transfer latencies and power consumption. FD sensing ensures a highly reliable connection.

In future work, we intend to extend our simulations to model end-to-end AR use cases by incorporating the cellular link between the companion device and the base station. A more rigorous quantitative analysis will help guide industry-level product development. A theoretical framework to analyze SL performance for AR is a promising research direction.

\section*{Acknowledgements}

This work was performed at Meta Platforms, Inc. when Ashutosh Srivastava was an intern at the company. The work was also supported in part by the NY State Center for Advanced Technology in Telecommunications (CATT), NYU Wireless, and the National Science Foundation (NSF) through RINGS: Resilient Edge Networks with Data-Driven Model-Based Learning under Grant CNS-2148309.
The authors would like to thank Dong Zheng, Neelakantan Krishnan, and Xiaodi Zhang for their valuable input. Authors Qing Zhao and Yee-Sin Chan were affiliated with Meta during the time of this work, but are no longer with the company. 

\bibliographystyle{IEEEtran}
\bibliography{IEEEabrv,references}

\end{document}